\documentclass[letterpaper]{aa520}

\usepackage{txfonts}
\usepackage{natbib}
\usepackage{aalongtable}
\usepackage{graphicx}
\usepackage{graphics}

\begin{document}

\title{Geometrical Tests of Cosmological Models}
\subtitle{II. Calibration of Rotational Widths and Disc Scaling Relations}

\author{Am\'{e}lie Saintonge\inst{1,2} \and  Karen L. Masters\inst{3} 
\and Christian Marinoni\inst{4}  \and Kristine Spekkens\inst{5} 
\and \\ Riccardo Giovanelli\inst{1} \and Martha P. Haynes\inst{1}}

\offprints{A. Saintonge, \email{amelie@physik.uzh.ch}}

\institute{
Department of Astronomy, Cornell University, Ithaca, NY 14853, USA
\and Institute for Theoretical Physics, University of Zurich, Winterthurerstrasse 190, CH-8057, Zurich, Switzerland
\and Harvard-Smithsonian Center for Astrophysics, Cambridge, MA 02143, USA
\and Centre de Physique Th\'eorique \thanks{Centre de Physique Th\'eorique is UMR 6207 -
 ``Unit\'e Mixte de Recherche'' of CNRS and of the Universities ``de Provence'',
 ``de la M\'editerran\'ee'' and ``du Sud Toulon-Var''- Laboratory
 affiliated to FRUMAM (FR 2291).}, CNRS-Universit\'e de Provence, Case 907,
 F-13288 Marseille, France
\and Department of Physics and Astronomy, Rutgers, the State University of New Jersey, Piscataway, NJ 08854, USA
}


\date{Received ... / Accepted ...}

\abstract{
This series of papers is dedicated to a new technique to select galaxies that can act as standard rods and standard candles in order to perform geometrical tests on large samples of high redshift galaxies to constrain different cosmological parameter. The goals of this paper are (1) to compare different rotation indicators in order to 
understand the relation between rotation velocities extracted from observations of the H${\alpha} \lambda$6563\AA\ line and the [OII]$\lambda 3727$\AA\ line, and (2) to determine the scaling relations between physical size, surface brightness and magnitude of galaxies and their rotation velocity using the SFI++, a large catalog of nearby galaxies observed at I-band. A good correlation is observed between the rotation curve-derived velocities of the H${\alpha}$ and [OII] observations, as well as between those calculated from velocity histograms, justifying the direct comparison of velocities measured from H${\alpha}$ rotation curves in nearby galaxies and from [OII] line widths at higher redshifts. To provide calibration for the geometrical tests, we give expressions for the different scaling relations between properties of galaxies (size, surface brightness, magnitude) and their rotation speeds.  Apart from the Tully-Fisher relation, we derive the size-rotation velocity and surface brightness-rotation velocity relations with unprecedentedly small scatters. We show how the best size-rotation velocity relation is derived when size is estimated not from disc scale lengths but from the isophotal diameter $r_{23.5}$, once these have been corrected for inclination and extinction effects.}


\titlerunning{Geometrical Tests of Cosmological Models II.}
\authorrunning{Saintonge et al.}

\maketitle

\section{INTRODUCTION}

With the new generation of telescopes and instruments, such as multi-object spectrograph and large field of view cameras, it has become possible to perform large, deep redshift surveys.  Using these data, the structure and geometry of the Universe can be studied through new channels, namely geometrical tests such as the angular diameter test (angular size - redshift relation), the Hubble diagram (magnitude - redshift relation) and the Hubble test (count - redshift relation).  These tests are generally difficult, requiring differentiation between the natural evolution of galaxies and the effects of geometry.  Large redshift surveys ease the task in that respect.

To perform the geometrical tests mentioned above, a population of objects that can be tracked through redshift needs to be defined.  More specifically, the angular diameter test requires a standard rod to be identified, the angular size of which is then being measured over the redshift range of interest.  The objects taken to serve that purpose could be anything from galaxies \citep[e.g.][]{sandage72}, clusters of galaxies \citep[e.g.][]{sereno,cooray98}, or dark matter halos \citep[]{cooray01}.  Regardless of its nature, a standard rod should satisfy simple criteria: its size should be measured reliably up to high redshifts, and it should be observable in the local Universe to provide calibration.  In this study, the relation between the physical size of a spiral galaxy and its rotation velocity (or global profile width) is used.  By virtue of this relation, selecting a population of objects with a given rotation velocity is equivalent to selecting them based on their physical size.  

In other papers of this series \citep[][thereafter Paper I and Paper III]{paperI,paperIII}, the angular diameter test is performed on a dataset of precursor Vimos/VLT Deep Survey (VVDS) \citep[]{vvds} galaxies, using the linewidth-diameter relation to select standard rods.  The test is used to isolate the effects of disk evolution with redshift, and to constrain the values of cosmological parameters.  Though the VVDS provides excellent data at high redshift, the angular diameter test highlights one of its shortcomings: limited volume coverage at low redshift and therefore the lack of a comparison sample in the local Universe.  

In this context, the goal of this paper is twofold.  First, using a large sample of nearby galaxies \citep[SFI++;][]{spr07} the physical properties of the standard rods and standard candles used to perform the geometrical tests are established, free from any evolutionary biases.  This is especially useful to calibrating the angular diameter relation and to separating the effects of galaxy evolution from those of geometrical variations (see Paper I).  By establishing clearly the scaling relations between the size, surface brightness and magnitude of these nearby galaxies observed at I-band and their rotation velocity, we establish the calibration that will allow us to select standard rods/candles simply by tracing through redshift galaxies with a given rotation speed. Having a strong handle on the structural parameters of disc galaxies in the local universe gives the unique opportunity of tracing over time the evolution of these properties for disc galaxies hosted in dark matter halos of the same mass, if rotation velocity is used as a proxy for halo mass, and compare them with predictions made under the hierarchical scenario for the growth of structures \citep[e.g.][]{mo98}.

The second goal of this paper is to compare the rotation velocity indicator used for the high redshift data to those used in the local Universe, cross-calibrating rotation indicators used at different redshifts. This last point is of relevance since rotation information for spiral galaxies can be obtained through the observation of various spectral lines, the choice of which varies with the redshift of the sample.  For galaxies with $z \lesssim 0.1$, the HI 21 cm line is an excellent candidate.  The H${\alpha} \lambda$6563\AA\ line is also frequently used for galaxies with low to moderate redshifts.  However, the H${\alpha}$ line is quickly redshifted into the near-infrared and becomes unavailable to ground observers using optical telescoptes for galaxies with $z \gtrsim 0.4$, even though with new near-infrared spectrographs such as SINFONI it is now possible to obtain H$\alpha$ rotation curves for high redshift galaxies \citep[e.g.][]{forster06}.  Even with the advent of such instruments, most large studies of galaxies at high redshift rely on the [OII]$\lambda 3727$ \AA\ line, including the VVDS and the DEEP2 Redshift Survey \citep[]{davis01}.  In order to compare sets of local and distant galaxies, it is therefore necessary to understand how rotation velocities extracted from these different lines relate.  It has already been shown that velocity widths derived from HI 21cm and H${\alpha}$ observations are in excellent agreement \citep[see for example][]{courteau97,vogt}.  The correlation between the [OII] and HI line widths has also been investigated \citep[]{kobulnicky}.  Using a sample of 22 nearby late-type spirals they find [OII] widths to be accurate to within $10 \%$ for galaxies with a roation width of 200 km s$^{-1}$, which is comparable to the overall scatter in the local Tully-Fisher relation.  However, they conclude that the uncertainties go up to about 50\% for galaxies with widths $<150$ km s$^{-1}$.  Here, a sample of 32 spiral galaxies with $0.155<z<0.25$ is used to compare velocity widths from the H${\alpha} \lambda$6563\AA\ and [OII]$\lambda 3727$ \AA\ lines (hereafter, H${\alpha}$ and [OII]).  Even though these galaxies are more distant than in the \citet[]{kobulnicky} sample, the scatter in the rotation curves data points is significantly smaller, resulting in a better estimation of the rotation velocities.   Our data also allows for a direct comparison between H${\alpha}$ and [OII], therefore bridging nicely the gap between low and high redshift samples.  Contrary to the \citet[]{kobulnicky} study, the galaxy sample used here is restricted to late spiral types, therefore we will not address the question of possible biasses due to varying morphologies.  

In \S \ref{data} a description of the galaxies observed and of the data reduction process is given, while in \S \ref{widths} we describe how velocity widths were extracted from the data and how they compare for both sets of emission lines.  In \S \ref{localtf} we presente a sample of nearby galaxies used as calibrators for the purpose of the angular diameter test (Paper III) and discussion of our results are given in \S \ref{discussion}.  Calculations were done assuming $H_0=70$\ km s$^{-1}$ Mpc$^{-1}$, $\Omega_M=0.3$, $\Omega_{\Lambda}=0.7$.

\section{DATA \label{data}}
\subsection{Target Selection and Observational Setup}
Using the Hale 5m telescope of Mount Palomar during the course of three short observing runs between 2003 March and 2004 February (total of 12 nights), long-slit spectra were obtained for a sample of spiral galaxies.  The galaxies were chosen in the area covered by the Data Release 2 of the {\it Sloan Digital Sky Survey}\footnotemark \ \citep[SDSS,][]{stoughton}, and selected to have inclinations in excess of 50 degrees, Petrosian radii containing $90\%$ of the light in r band larger than 4\arcsec \ to avoid AGN-dominated systems, and [OII]$\lambda3727$\AA \ equivalent widths in the range of 10-35\AA.\  The galaxies were also required to be in the redshift range $0.155<z<0.25$.  Galaxies with close neighbours or disturbed morphologies were rejected to avoid including interacting systems in the sample.

Using the Double Spectrograph \citep[]{okegunn}, both H${\alpha}$ and [OII] emission lines could be observed simultaneously.  The main properties of the galaxies are listed in Table \ref{galaxies}.  The first two columns are the AGC ({\it Arecibo General Catalog}) and SDSS names, respectively (the J2000 coordinates of the objects can be directly extracted from their SDSS name).  The third column is the heliocentric systemic velocity of the galaxies, as measured from the H${\alpha}$ line. All other parameters (redshift, I-band magnitude, disc scale-length, inclination and position angle measured counterclockwise from North) are from the SDSS database. Unfortunately, about 50\% of the observing time was lost to bad weather conditions, forcing the final sample presented in this paper to contain only the 32 galaxies for which it was possible to get rotation curves in both H${\alpha}$ and [OII] . 

\footnotetext{Funding for the creation and distribution of the SDSS Archive has been provided by the Alfred P. Sloan Foundation, the Participating Institutions, the National Aeronautics and Space Administration, the National Science Foundation, the U.S. Department of Energy, the Japanese Monbukagakusho, and the Max Planck Society. The SDSS Web site is http://www.sdss.org/.
The SDSS is managed by the Astrophysical Research Consortium (ARC) for the Participating Institutions. The Participating Institutions are The University of Chicago, Fermilab, the Institute for Advanced Study, the Japan Participation Group, The Johns Hopkins University, Los Alamos National Laboratory, the Max-Planck-Institute for Astronomy (MPIA), the Max-Planck-Institute for Astrophysics (MPA), New Mexico State University, University of Pittsburgh, Princeton University, the United States Naval Observatory, and the University of Washington.}

The red camera of the Double Spectrograph was used with a 1200 lines mm$^{-1}$ grating (7100\AA\ blaze) set at an angle of 47\degr 18\arcmin\ in order to cover the wavelength range of 7520-8185 \AA.  The spectral resolution is 0.65 \AA\ pixel$^{-1}$ with 24$\mu$m pixels and the spatial scale is of 0\farcs468 pixel$^{-1}$. The CCD camera had a size of 1024$\times$1024 pixels, and a 1\arcsec$\times$128\arcsec\ slit was used (except for a few galaxies where the 2\arcsec$\times$128\arcsec\ slit was used due to very poor seeing conditions).  The blue camera was set to cover the range 3675-5140 \AA\ by using a 1200 lines mm$^{-1}$ grating with a 4700\AA\ blaze at an angle of 35\degr 52\arcmin.  Using the 2788$\times$512 chip (15$\mu$m pixels), a spatial sampling of 0\farcs390 pixel$^{-1}$ was achieved, as well as a 0.55 \AA\ pixel$^{-1}$ spectral resolution.

The exposure time was of 3600 s for most galaxies, though it was extended to 4800 s for a few sources when the flux in the [OII] line was very low.  Sky transparency and seeing conditions varied greatly over the course of the observing runs, but most observations were made between 1\arcsec\ and 2\arcsec\ seeing.  In order to align the slit with the major axis of the galaxies, the ring angle of the spectrograph was set using the position angles of the galaxies given in the SDSS database.  For each galaxy, the ring angle used is given in Table \ref{galaxies}.  On each night of observing, standard calibration images were obtained: dome flats, twilight flats (for the blue camera only), bias frames and darks.  Given observing conditions, no attempt was made for any photometric calibration.

\begin{table*}
\caption{Properties of the Galaxy Sample}
\label{galaxies}
\centering
\begin{tabular}{lccccccc}
\hline\hline
AGC \# & Name & $v_{0}$ & $z_{SDSS}$ & $m_i$ & $r_D$ & $i$ & PA \\
 & & km\.s$^-1$ & & mag & kpc & deg & deg\\
\hline
101738 &\object{SDSS J001107.7+003551.7} &46600 &0.155 &17.81 &4.24 &66 &8\\
101701 &\object{SDSS J003123.8+141313.8} &49444 &0.165 &17.15 &6.93 &67 &116\\
112393 &\object{SDSS J014451.5+133616.9} &46711 &0.156 &17.31 &6.18 &54 &111\\
130796 &\object{SDSS J031818.8+010750.4} &48640 &0.162 &17.68 &3.29 &51 &87\\
431415 &\object{SDSS J032434.8-071050.6} &49673 &0.166 &17.32 &7.82 &64 &356\\
130826 &\object{SDSS J033331.8+010717.1} &53337 &0.178 &17.48 &6.92 &61 &158\\
431418 &\object{SDSS J040000.2-053852.3} &50138 &0.167 &17.36 &7.03 &66 &51\\
171385 &\object{SDSS J073235.0+374734.4} &65413 &0.219 &16.94 &9.09 &62 &331\\
171389 &\object{SDSS J074608.6+395658.5} &47142 &0.157 &17.46 &5.54 &49 &343\\
171394 &\object{SDSS J075555.0+364057.7} &55351 &0.185 &17.01 &7.60 &49 &103\\
181489 &\object{SDSS J080438.6+463536.6} &60671 &0.202 &17.12 &6.00 &55 &12\\
181487 &\object{SDSS J081345.7+473552.5} &54478 &0.182 &17.40 &6.28 &61 &322\\
181490 &\object{SDSS J081750.4+481511.3} &55617 &0.186 &17.30 &7.82 &56 &35\\
181488 &\object{SDSS J084304.2-000737.6} &49300 &0.165 &17.40 &8.79 &63 &7\\
191728 &\object{SDSS J091957.0+013851.6} &52850 &0.176 &16.90 &5.51 &51 &20\\
191729 &\object{SDSS J092050.9+575758.5} &51558 &0.172 &16.94 &6.01 &46 &117\\
191731 &\object{SDSS J094412.5+020206.5} &47453 &0.158 &17.17 &6.30 &60 &99\\
191732 &\object{SDSS J100047.6+020406.6} &55778 &0.185 &16.71 &5.02 &61 &90\\
501752 &\object{SDSS J100757.3-005612.3} &46402 &0.155 &17.08 &4.38 &58 &179\\
202041 &\object{SDSS J104215.5+010508.3} &58452 &0.195 &17.46 &7.47 &60 &104\\
501753 &\object{SDSS J110117.7-010014.9} &47277 &0.158 &17.39 &7.15 &66 &36\\
212861 &\object{SDSS J110936.9+010957.8} &63918 &0.213 &17.42 &9.08 &60 &95\\
511373 &\object{SDSS J114001.2-004624.3} &56368 &0.188 &17.36 &5.25 &58 &107\\
212862 &\object{SDSS J114919.0+001806.3} &48216 &0.161 &17.14 &6.40 &53 &102\\
511374 &\object{SDSS J115640.0-005128.6} &50585 &0.169 &17.13 &6.66 &55 &122\\
521390 &\object{SDSS J123824.5-013339.8} &56357 &0.189 &17.03 &6.06 &54 &41\\
224090 &\object{SDSS J124819.3+672312.2} &50790 &0.170 &17.38 &7.56 &51 &26\\
224092 &\object{SDSS J125522.8+672306.8} &49401 &0.165 &17.51 &6.73 &55 &173\\
242064 &\object{SDSS J141054.1+013714.3} &51706 &0.173 &17.32 &8.46 &49 &129\\
242065 &\object{SDSS J141322.7+644404.7} &54763 &0.183 &17.51 &5.04 &53 &79\\
242067 &\object{SDSS J142846.2-000245.6} &55232 &0.184 &17.23 &7.75 &58 &242\\
332257 &\object{SDSS J234325.9+005605.2} &67105 &0.224 &18.20 &7.45 &64 &51\\
\hline
\end{tabular}
\end{table*}

\subsection{Data Reduction}
The data from both cameras were reduced independently using tasks from the Image Reduction and Analysis Facility (IRAF) \citep[]{iraf}.  Images from the blue camera had to be reduced with particular caution due to the presence of many bad columns and other defects in the CCD.  First, all the images were trimmed to remove overscans and low sensitivity regions on the edges.  Then flat fields were constructed separately for each camera and for each night using the dome flats, as well as the twilight flats in the case of the blue images.  After this, the images were also corrected for bad pixels using a mask created from twilight flats and the effects of cosmic rays removed.

The cleaned images were wavelength calibrated and corrected for s-shape distortion.  This was done in the dispersion direction by using observations of a reference star that was positioned at five different places along the slit, and for the spatial direction by using either strong sky lines (red camera images) or lamp spectra (blue camera images).  The correction was in some cases important, especially in the dispersion direction where it was generally on the order of a few pixels.  Individual exposures of each galaxy were then appropriately shifted and stacked.  For these combined images, the background was fitted in order to remove the sky lines.  The process allowed for optimal removal of the sky lines falling near or on the galaxy emission lines of interest.  The continuum was finally subtracted from these background corrected frames.

\section{COMPARING AND CALIBRATING ROTATION VELOCITY INDICATORS \label{widths}}

Using these reduced data, it was possible to extract rotation velocity information for the galaxies in this Palomar sample (hereafter referred to as the $z0.2$ sample) using both H$\alpha$ and [OII] emission lines and two different techniques: rotation curves and velocity histograms. We here describe these two different techniques.

\subsection{Rotation Curve Extraction and Modeling}
A Gaussian was fitted along the dispersion axis at every pixel inside a box encompassing the emission line (either H${\alpha}$ or [OII]).  In the case of the [OII] line where the signal-to-noise ratio (S/N) was often smaller than 3, Hanning smoothing was performed before the Gaussian fitting took place.  Then, the rotation curves were calibrated, the measured wavelength converted to a velocity and the radial axis scaled to arcseconds.  Following this, each point of the fit was examined and either rejected or included in the rotation curve.  The criteria were a minimal S/N for the fitted Gaussian (4 for H${\alpha}$ and 3 for [OII]) and a maximum width of 400 km s$^{-1}$.  The resulting rotation curves were then examined and some rejected points reintroduced, while some others that satisfied the criteria were judged unrealistic and manually removed.  These points for which a manual intervention was required were almost all either in the outer regions of the galaxy where the S/N drops significantly or contaminated by sky lines or cosmic rays that were not completely successfully removed.  Finally, the center of light of the galaxy was determined and used to center the rotation curve along the spatial direction.  This procedure is similar to the one presented in \citet[]{vogt}.

\begin{figure*}
\centering
  \includegraphics[width=17cm]{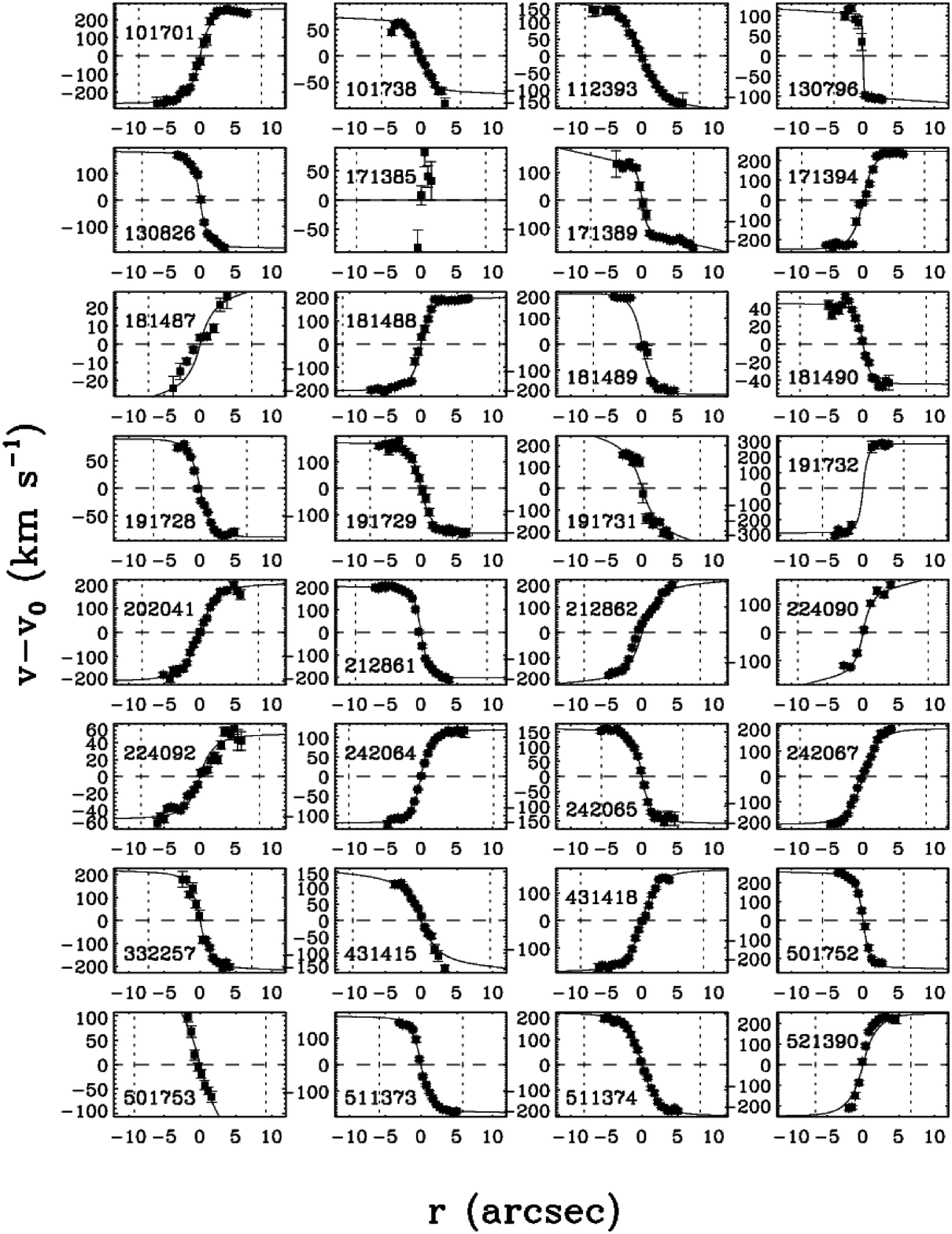}
  \caption{The H${\alpha}$ rotation curves and their Polyex fits for the 32 spirals of the $z0.2$. The dotted lines show for each galaxy the position of $r_{83}$, the radius containing 83\% of the light, and the solid line is the Polyex fit (when sufficient data was available).}
  \label{rcHa}
\end{figure*}

\begin{figure*}
\centering
  \includegraphics[width=17cm]{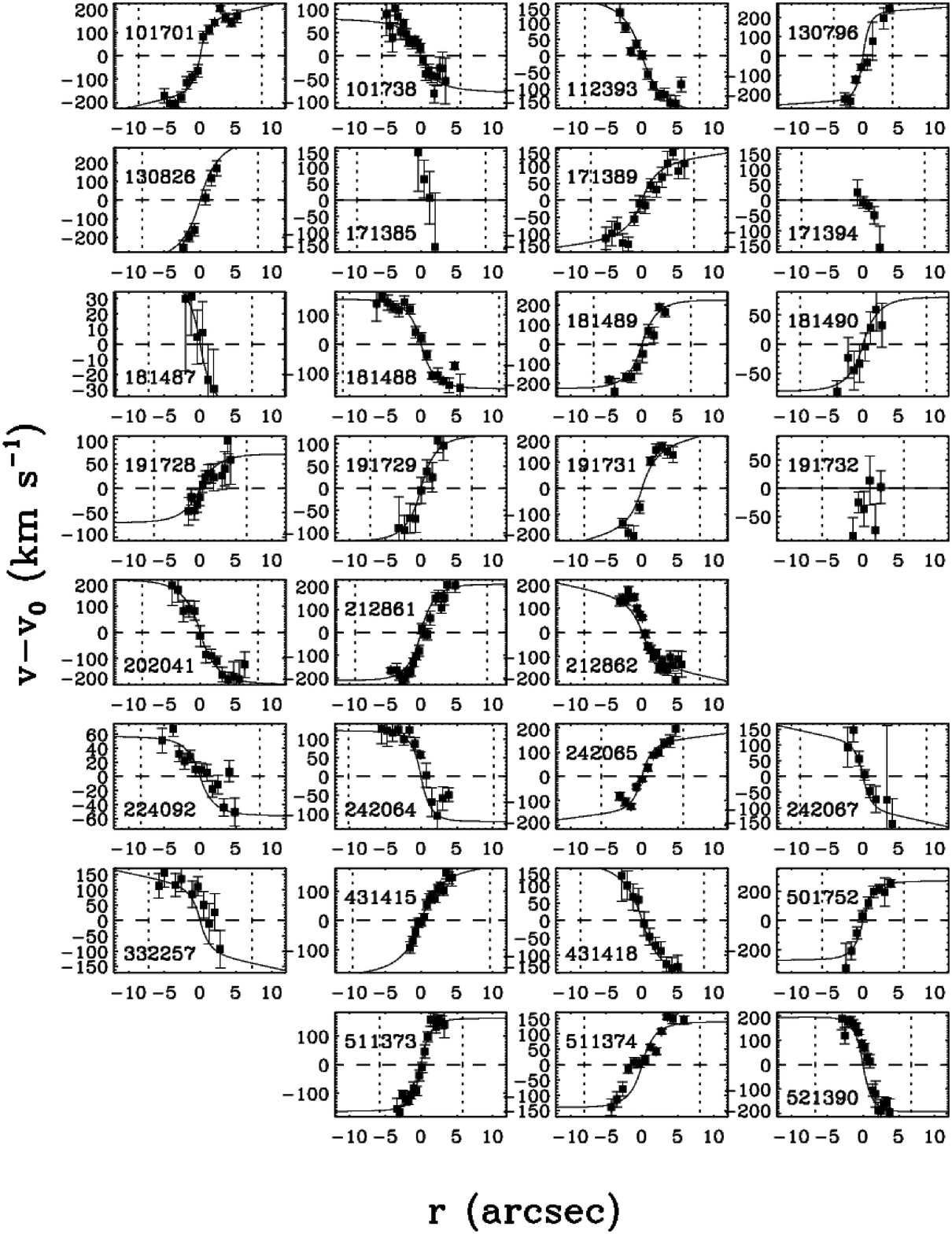}
  \caption{The [OII] rotation curves and their Polyex fits for the galaxies which had detectable [OII]$\lambda3727\AA$ emission. The solid line represents the Polyex fit (when sufficient data was available), and the dotted lines show for each galaxy the position of $r_{83}$, the radius containing 83\% of the light.} 
  \label{Orc}
\end{figure*}

The rotation curves were subsequently modeled by first folding about the kinematic center and fitting a Polyex model \citep[]{polyex} of the form:
\begin{equation}
V(r)=a_1(1-e^{-a_2r})(1+a_3r).
\end{equation}
The parameter $a_1$ is related to the maximal velocity of the rotation curve and $a_2$ controls the exponential rise of the rotation curve and the position of the ``elbow'', the point where a transition occurs between the steeply rising inner part and the flat outer region.  Finally, $a_3$ controls the slope of the outer part of the rotation curve.  Due to the small radial extent of the emission lines, the value of $a_3$ is meaningless in most cases and has been manually set to zero.  The H${\alpha}$ rotation curves are shown for all galaxies in Figure \ref{rcHa}, and the [OII] rotation curves in Figures \ref{Orc}.  The solid lines are the Polyex fits, while the vertical dotted lines show the position of $r_{83}$, the radius containing $83\%$ of the light of the galaxy.

\begin{figure*}
\centering
  \includegraphics[width=17cm]{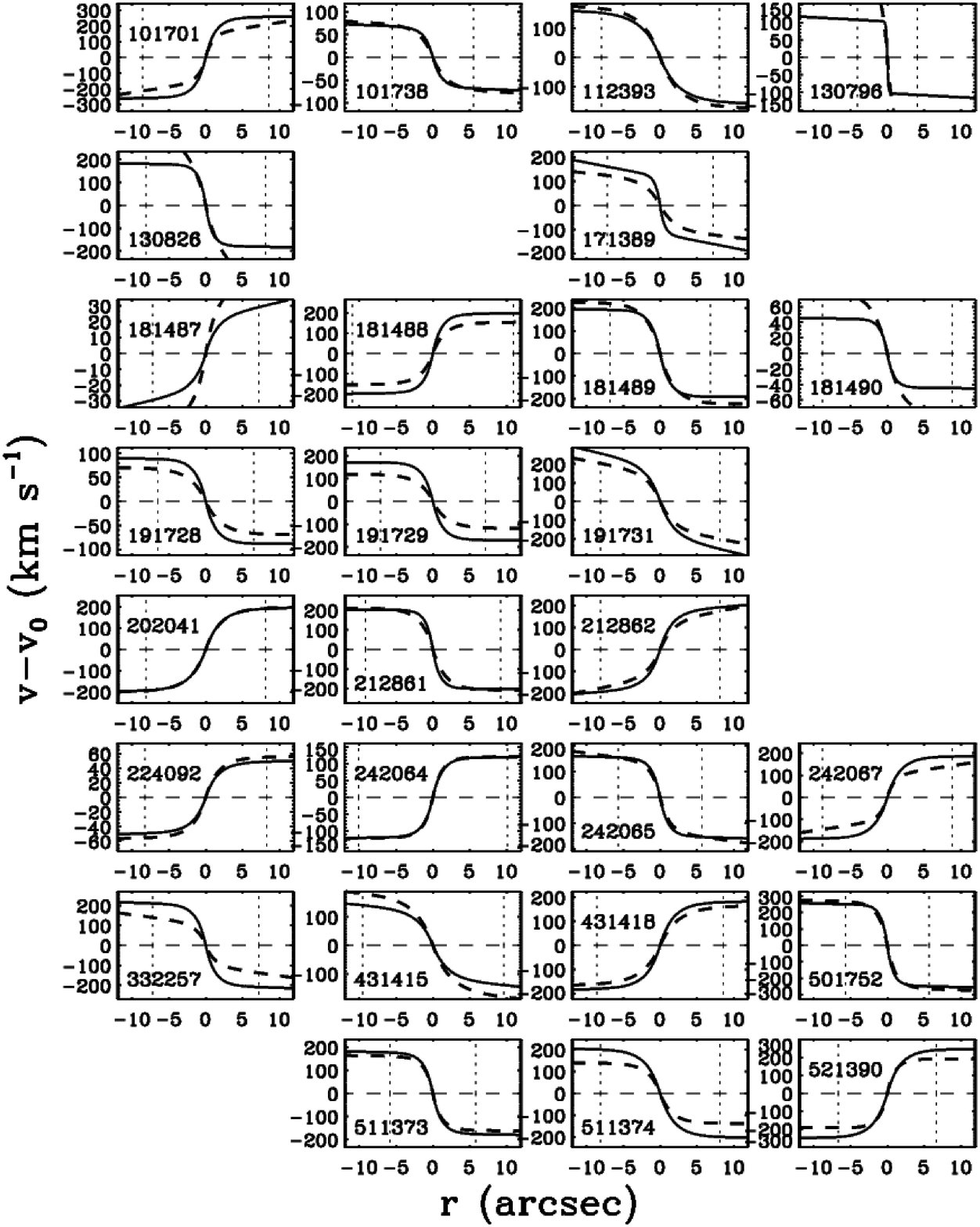}
  \caption{Polyex fits to the H$\alpha$ (solid line) and to the [OII] (dashed line) rotation curves presented in Figures \ref{rcHa} and \ref{Orc}, when both fits are available.  Again, the vertical dotted lines show for each galaxy the position of $r_{83}$.} 
  \label{comp_polyex}
\end{figure*}

In Figure \ref{comp_polyex}, the Polyex fits to both the H$\alpha$ and the [OII] rotations curves are shown together when sufficient data was available in both spectral lines to allow for the fitting process. In all the cases where the [OII] line was bright enough for the rotation curve to be traced out to the region where it starts flattening out, the agreement between the two sets of rotation curves is excellent. In the other cases, the trend is for the [OII] rotation curves to be slightly shallower.

\begin{table*}
\caption{Galactic Rotation Properties, from the H${\alpha}$ Emission Line}
\label{tabHa}
\centering
\begin{tabular}{lccccc}
\hline\hline
AGC \# & $a_1$ & $a_2$ & $a_3$ & $v_{rot}(r_{83})$ & $\sigma_{rot}$\\
& km s$^{-1}$ & (asec$^{-1}$) & (asec$^{-1}$) & (km s$^{-1}$) & (km s$^{-1}$\\
\hline
101701 &253.4 &0.7800 & 0.00255 &$258.6\pm18.3$ &$275.6\pm12.5$\\
101738 & 65.1 &0.9494 & 0.01000 &$ 68.3\pm10.8$ &$ 92.9\pm 3.5$\\
112393 &151.6 &0.4260 & 0.00436 &$151.8\pm 6.7$ &$144.5\pm 6.9$\\
130796 &102.9 &9.2192 & 0.01120 &$107.6\pm17.5$ &$181.2\pm12.6$\\
130826 &177.4 &1.1939 & 0.00319 &$182.0\pm10.0$ &$197.6\pm11.3$\\
171385 &...  &... &... &... &...\\
171389 &118.1 &1.8150 & 0.05000 &$160.3\pm10.7$ &$179.4\pm20.4$\\
171394 &249.5 &0.7820 &-0.00029 &$248.6\pm25.4$ &$236.8\pm25.4$\\
181487 & 21.7 &0.7317 & 0.05000 &$ 29.3\pm 6.0$ &$ 62.0\pm 3.5$\\
181488 &192.9 &0.9066 & 0.00285 &$198.8\pm12.0$ &$200.6\pm22.9$\\
181489 &194.3 &0.8190 &-0.00054 &$192.8\pm20.6$ &$175.9\pm 8.7$\\
181490 & 44.1 &1.1869 & 0.00159 &$ 44.7\pm 5.3$ &$ 78.7\pm 3.4$\\
191728 & 89.2 &0.7410 &-0.00040 &$ 88.2\pm 7.2$ &$ 73.0\pm 3.0$\\
191729 &170.9 &0.8057 & 0.00043 &$170.9\pm11.4$ &$236.5\pm11.9$\\
191731 &183.6 &0.6494 & 0.05000 &$256.3\pm39.9$ &$217.2\pm50.4$\\
191732 &282.4 &1.6846 & 0.00016 &$282.6\pm14.1$ &$223.9\pm50.4$\\
202041 &193.0 &0.5153 & 0.00204 &$193.1\pm13.2$ &$202.8\pm12.0$\\
212861 &201.7 &1.1360 & 0.00037 &$202.4\pm 9.7$ &$199.1\pm10.8$\\
212862 &173.8 &0.6494 & 0.01500 &$193.8\pm21.9$ &$144.3\pm 9.9$\\
224090 &131.0 &1.0978 & 0.05000 &$190.7\pm12.9$ &$203.7\pm10.7$\\
224092 & 47.5 &0.6303 & 0.00505 &$ 49.3\pm 7.9$ &$ 77.9\pm 5.9$\\
242064 &117.3 &0.8463 & 0.00211 &$119.8\pm 3.6$ &$131.6\pm 7.0$\\
242065 &150.9 &0.9871 & 0.00494 &$154.6\pm 8.1$ &$151.8\pm 5.6$\\
242067 &186.1 &0.5976 & 0.00119 &$187.0\pm16.0$ &$131.4\pm 5.6$\\
332257 &202.6 &0.8316 & 0.00573 &$210.5\pm15.8$ &$192.6\pm18.2$\\
431415 &119.2 &0.5455 & 0.02000 &$141.5\pm14.7$ &$ 84.3\pm13.7$\\
431418 &176.3 &0.5846 & 0.00356 &$180.5\pm10.8$ &$152.8\pm 4.8$\\
501752 &243.9 &1.3028 & 0.00417 &$249.5\pm 6.2$ &$245.5\pm11.3$\\
501753 &186.1 &0.3137 & 0.00399 &$182.0\pm12.3$ &$106.6\pm35.1$\\
511373 &175.2 &0.9155 & 0.00283 &$177.2\pm 8.1$ &$149.6\pm 6.0$\\
511374 &199.5 &0.5625 & 0.00185 &$200.3\pm 7.9$ &$187.9\pm 6.8$\\
521390 &248.8 &0.5854 & 0.00000 &$243.9\pm30.8$ &$251.0\pm 8.3$\\
\hline
\end{tabular}
\end{table*}

\begin{table*}
\caption{Galactic Rotation Properties, from the [OII] Emission Line}
\label{tabOII}
\centering
\begin{tabular}{lccccc}
\hline\hline
AGC \# & $a_1$ & $a_2$ & $a_3$ & $v_{rot}(r_{83})$ & $\sigma_{rot}$\\
& km s$^{-1}$ & (asec$^{-1}$) & (asec$^{-1}$) & (km s$^{-1}$) & (km s$^{-1}$)\\
\hline
101701 &148.9 &1.1659 & 0.05000 &$212.8\pm23.5$ &$217.0\pm 53.3$\\
101738 & 70.5 &0.5846 & 0.01000 &$ 71.5\pm18.4$ &$ 84.7\pm 13.2$\\
112393 &166.7 &0.4304 & 0.00436 &$167.1\pm 9.1$ &$186.4\pm 56.6$\\
130796 &226.2 &1.2712 & 0.01120 &$235.4\pm45.9$ &$260.1\pm 47.1$\\
130826 &318.4 &0.4384 & 0.00319 &$317.0\pm13.5$ &$185.2\pm111.4$\\
171385 &...  &... &... &... &...\\
171389 &103.3 &0.5794 & 0.03000 &$123.5\pm24.2$ &$205.8\pm 73.7$\\
171394 &... &... &... &... &$283.8\pm 86.2$\\
181487 & 42.7 &0.6560 & 0.00288 &$ 43.2\pm 5.8$ &$121.5\pm 41.7$\\
181488 &150.0 &0.6544 & 0.00285 &$154.5\pm15.2$ &$212.4\pm 52.2$\\
181489 &226.5 &0.5540 &-0.00054 &$220.3\pm35.9$ &$296.2\pm 83.2$\\
181490 & 79.5 &0.5449 & 0.00159 &$ 79.9\pm11.7$ &$116.6\pm 54.6$\\
191728 & 70.5 &0.4750 &-0.00040 &$ 67.1\pm16.5$ &$176.6\pm 31.2$\\
191729 &118.7 &0.5453 & 0.00043 &$116.6\pm18.6$ &$271.8\pm 78.3$\\
191731 &145.2 &0.6494 & 0.05000 &$202.7\pm43.6$ &$194.6\pm 59.1$\\
191732 &... &... &... &... &$279.1\pm 72.9$\\
202041 &195.2 &0.4955 & 0.00204 &$194.8\pm14.0$ &$225.3\pm 60.8$\\
212861 &209.5 &0.6183 & 0.00037 &$209.5\pm24.4$ &$129.9\pm 37.8$\\
212862 &126.6 &0.6494 & 0.05000 &$176.8\pm31.1$ &$155.4\pm 34.5$\\
224090 &...  &... &... &... &...\\
224092 & 53.5 &0.6303 & 0.00505 &$ 55.5\pm13.6$ &$117.6\pm 34.1$\\
242064 &119.2 &0.8766 & 0.00211 &$121.7\pm15.7$ &$148.2\pm 44.7$\\
242065 &131.8 &0.9146 & 0.03000 &$153.6\pm28.1$ &$109.1\pm 20.9$\\
242067 &100.9 &1.1304 & 0.05000 &$145.1\pm29.0$ &$166.3\pm 76.6$\\
332257 &102.5 &0.9018 & 0.05000 &$139.4\pm19.1$ &$141.5\pm 66.0$\\
431415 &161.7 &0.4619 & 0.01500 &$182.9\pm 6.6$ &$ 92.6\pm 29.5$\\
431418 &159.4 &0.4397 & 0.00356 &$160.5\pm 9.0$ &$229.1\pm 43.2$\\
501752 &260.5 &0.8905 & 0.00417 &$265.0\pm40.8$ &$369.2\pm145.5$\\
501753 &137.1 &0.7214 & 0.00399 &$142.0\pm 5.6$ &$175.7\pm 49.0$\\
511373 &158.9 &0.9331 & 0.00283 &$160.8\pm13.7$ &$189.7\pm 30.7$\\
511374 &136.8 &0.6494 & 0.00185 &$138.1\pm29.6$ &$182.7\pm 28.7$\\
521390 &194.1 &0.9440 &-0.00068 &$192.9\pm11.7$ &$307.2\pm 53.1$\\
\hline
\end{tabular}
\end{table*}

Using these fits, the rotational velocity of each galaxy was determined, both for its H${\alpha}$ and [OII] lines.  The velocity adopted for a given rotation curve is the value of the Polyex fit at a radius corresponding to $r_{83}$.  The values for $r_{83}$ are taken to be $3.5\times r_D$, where $r_D$ is the disk scale length.  Since all the galaxies in the sample were taken from the SDSS, the values of physical parameters were taken from their database.  The scale lengths are from an exponential fit performed on the i-band image of each galaxy.  These values are presented in column 5 of Table \ref{galaxies}.  The value of $r_{83}$ for each galaxy is also represented in Figures \ref{rcHa} and \ref{Orc} by the vertical dotted lines.  Also, Tables \ref{tabHa} and \ref{tabOII} present the values of the Polyex coefficients and the rotation velocity at $r_{83}$, $v_{rot}(r_{83})$, for the H${\alpha}$ and [OII] data, respectively.

\begin{figure}
\resizebox{\hsize}{!}{\includegraphics{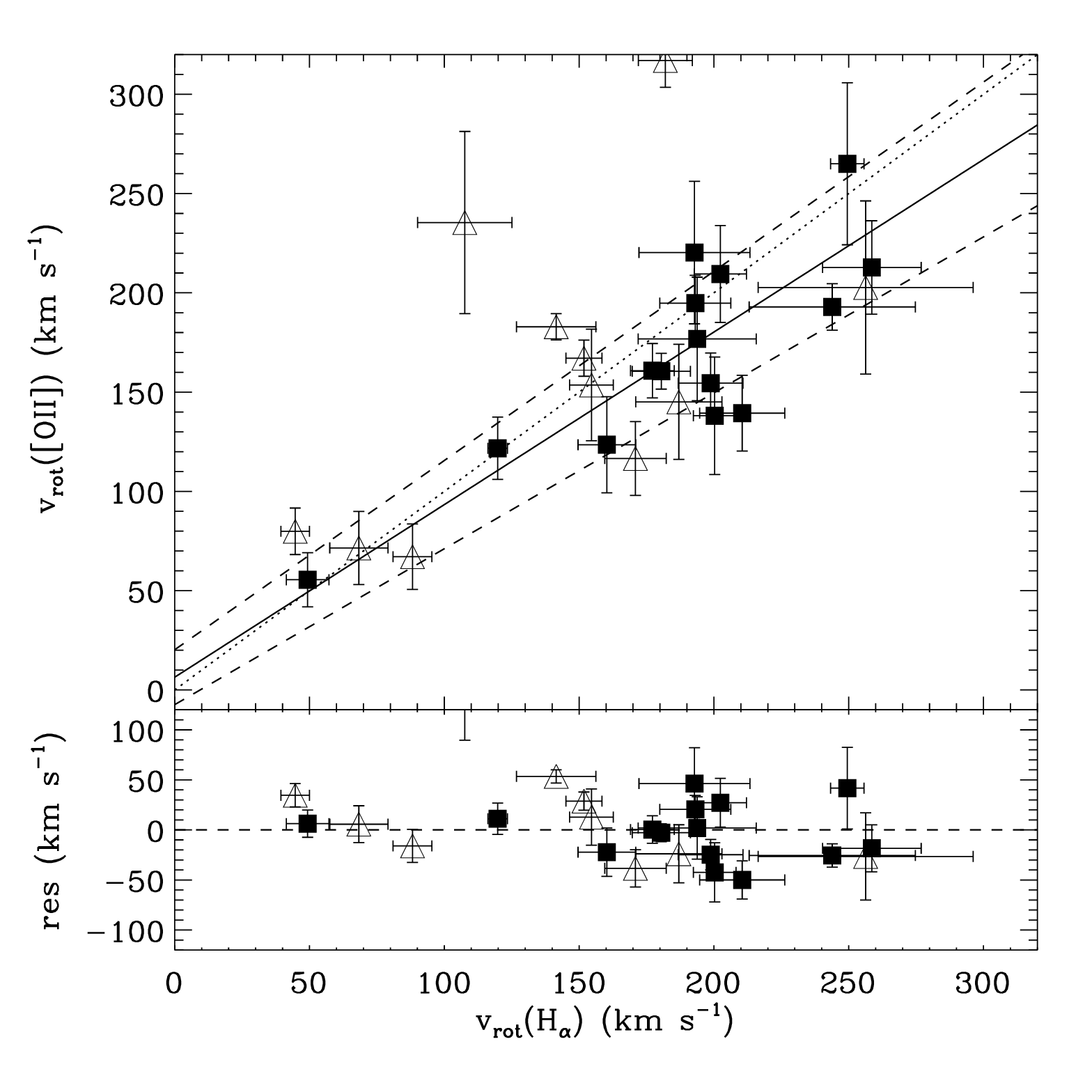}}
\caption{Comparison of the rotational velocities derived from the rotation curves for H${\alpha}$ and [OII].  {\it Top Panel:} the solid line is the best fit obtained by an ordinary least-squares regression, excluding the two outlying points.  The dashed lines are the maximum and minimum slopes determined by the errors on the parameters of the best-fit linear relation and the dotted line corresponds to a 1:1 relation. {\it Bottom Panel:} the residuals of the best-fit model.  In both panels, the filled symbols represent the 15 galaxies with the most secure rotation velocity estimates, while the open symbols are those galaxies with poorer rotation curve extent, either in H${\alpha}$ or [OII].}
\label{vrot_plot}
\end{figure}

In Figure \ref{vrot_plot}, we compare the rotation velocities that come from the H${\alpha}$ and [OII] rotation curves.  Following the prescription of \citet[]{isobe}, a best fit to the data was obtained by linear regression using an ordinary least-squares bisector method (also called ``double regression'').  The two outlying data points were ignored during the fitting.  The solid line in Figure \ref{vrot_plot} is the best-fit relation and the two dashed lines are the minimum and maximum slopes allowed by the best-fit parameters.  The dotted line is the 1:1 relation.  The standard deviation of the residuals of the best-fit (shown in the bottom panel) is 28.6 km s$^{-1}$.

\subsection{Velocity Histograms}

In order to cross-correlate the results with techniques used at higher redshift, measurements of the rotation velocities using velocity histograms were also made.  These velocity histograms were built by collapsing the two-dimensional images along the spatial direction in order to form one-dimensional spectra.  To determine the region of the image that will be collapsed, the central position of the continuum along the spatial direction as well as its width were determined by fitting a simple Gaussian.  This region needs to be large enough to include all the emission from the galaxy, but not unnecessarily wide to avoid increasing the noise in the one-dimensional spectra.  Since some of the galaxies have their emission lines extending slightly beyond the continuum, everything within $5\sigma$ of the center of the continuum was selected and added together, thus producing an intensity weighted spectrum.

\begin{figure}
\resizebox{\hsize}{!}{\includegraphics{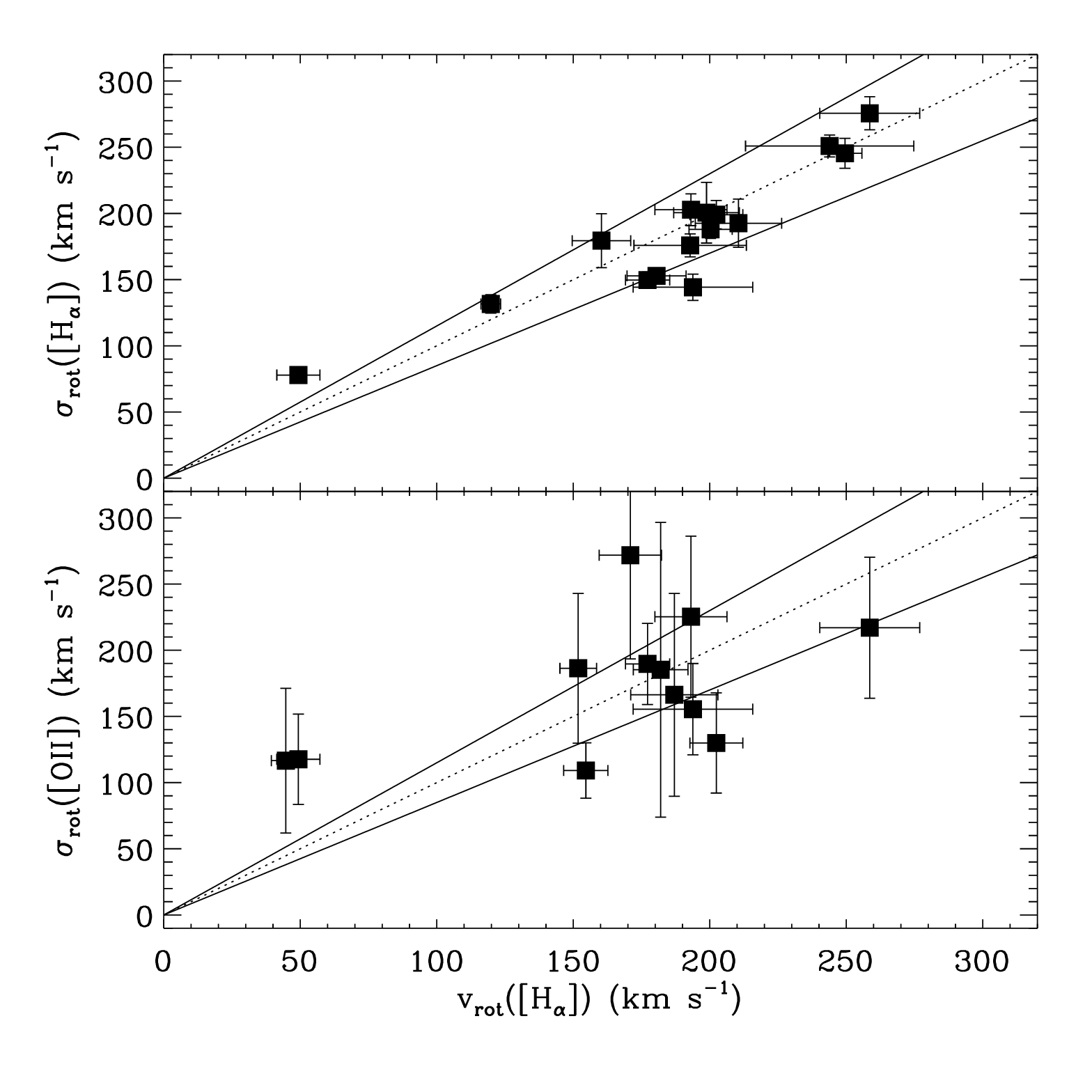}}
\caption{Comparison of the rotational velocities derived from the rotation curves ($v_{rot}$) and from the one-dimensional collapsed spectra ($\sigma_{rot}$) for H${\alpha}$ (top panel) and [OII] (bottom panel).  In each pannel the dotted line corresponds to a 1:1 relation, and the continuous lines a deviation of $15\%$ from it.}
\label{sigma_plot}
\end{figure}

A Gaussian was fitted to each of the H${\alpha}$ histograms.  The full width at half maximum (FWHM) of that Gaussian was converted into an estimate of the rotation velocity of the galaxy, $\sigma_{rot}$, via:
\begin{equation}
\sigma_{rot}=\frac{({\rm FWHM}_{l}^2-{\rm FWHM}_{i}^2)^{1/2}}{2.355} \cdot \frac{c}{\lambda_{0}(1+z)} \cdot \frac{1}{\sin(i)}
\label{sigma_eqn}
\end{equation}
where FWHM$_{l}$ is the FWHM of the emission line histogram, FWHM$_{i}$ the instrumental broadening, $\lambda_{0}$ the rest-frame wavelength of the emission line and $i$ the inclination of the disk.  The inclinations were determined from the axial ratio of the disks as derived from the same SDSS fits that produced the disk scale lengths.  In the case of the H${\alpha}$ lines, FWHM$_{i}$ was obtained by fitting a Gaussian to a bright night sky line that falls near the emission line.  The velocities obtained using this technique ($\sigma_{rot}$) are plotted against the velocities derived from the rotation curves ($v_{rot}$) for the H${\alpha}$ emission on the top panel of Figure \ref{sigma_plot} and are presented in column 6 of Table \ref{tabHa}.  In Figure \ref{sigma_plot} we plot only the galaxies for which the rotation curve extends radially far enough to reliably determine a rotation velocity.  The standard deviation of the residuals of the $\sigma_{rot}(H{\alpha})-v_{rot}(H{\alpha})$ from the 1:1 relation is 20.8 km s$^{-1}$, which represents an error of $\sim 10 \%$ for a galaxy with a rotation velocity of 200 km s$^{-1}$.

For the [OII] emission, one-dimensional spectra were produced in the same manner, but this time $\sigma_{rot}$ cannot be calculated by a single Gaussian fit since the doublet nature of the line needs to be taken into account.  Instead, deblending of the two lines was done using the ``splot'' procedure in IRAF.  The expected positions of the two lines of the doublet were given as input, and the relative position of the two lines is maintained fixed during the fitting process.  We farther required that the width of the two lines be the same (as it should be for physical reasons), and that the amplitude of the two lines be in a reasonable ratio (not more than a factor of three between the two, which is similar to the criteria of \citet[]{kobulnicky}).  Since the process relies significantly on subjective judgement on the part of the data processor, especially with regard to the size of the fitting region around the emission line, it was performed independently by two of us (AS, CM).  The results were compared and a quality flag was put on the values.  A width estimate was judged reliable when the two values agreed within the spectral resolution of the observations (i.e. 0.55\AA).  The widths were transformed into rotation velocities using eq. \ref{sigma_eqn}.  This time, however, FWHM$_{i}$ is calculated using exposures of a FeAr hollow cathode calibration lamp, since there are very few bright sky lines over the relevant wavelength range.  The calculated velocities ($\sigma_{rot}$) are plotted against the rotation curve derived velocities for H${\alpha}$ on the bottom panel of Figure \ref{sigma_plot} and given in the last column of Table \ref{tabOII}.  The rms of the residuals of the $\sigma_{rot}([OII])-v_{rot}(H{\alpha})$ relation is 52.7 km s$^{-1}$, which at any rotation width represents an error larger by a factor of $\sim2.5$ than the relation for H${\alpha}$ alone.

\subsection{Definition of the Rotation Classes \label{rot_classes}}
When comes the time to perform the geometrical tests, the VVDS rotation velocities are used to separate the galaxies into two samples, one of slow and one of fast rotators.  This is done in order to trace out the evolution of two different populations of galaxies by using the angular diameter test (see Paper I and Paper III). These populations are defined as the  ``slow rotators'', which have $0 < v_{rot} < 100$ km s$^{-1}$ and the ``fast rotators" with $100 < v_{rot} < 200$ km s$^{-1}$. The analysis of the rotation curves and of the velocity histograms for the $z0.2$ sample has shown that even considering the scatter in the relation between the different velocity indicators and the different spectral lines, similar criteria on rotation velocities can be applied across the different data sets to define the populations of objects that can act as standard rods or candles to perform the geometrical tests under the approach suggested in Paper I. Because of the restricted size of the $z0.2$ sample and the very small number of "slow rotators" with reliable velocity measurements, these galaxies will not be used directly to perform the geometrical tests in Paper III. However, given the bridge that we have just established between low and high redshift velocity indicators, one could eventually use large data sets of nearby galaxies, such as SDSS, to complete the gap in redshift coverage between the very local SFI++ catalog (see next section), and the VVDS data.

\section{CALIBRATING THE STANDARD RODS AND STANDARD CANDLES}

In order to perform some geometrical tests such as the angular diameter test, not only will we need the well-defined samples of galaxies at $0.15 \lesssim z \lesssim 1.4$ but also reference values for the local Universe ($z \approx 0$). Since the contributions of galaxy evolution and geometric signature are superposed, it is essential to determine the sizes and luminosities of the standard rods/candles for a local sample that is free from any evolution, and also the scatter in these relations.

\subsection{A Sample of Galaxies in the Local Universe \label{localtf}}

The SFI++ \citep{spr07}, a sample of $\sim 4200$ local galaxies, is used for this purpose.  They range in distance from 0 to 300 Mpc, with most objects within 100 Mpc.  SFI++ builds on the earlier SFI (Spiral Field I-band) and SCI (Spiral Cluster I-band) samples published in a series of papers in the 1990s (e.g. \citet{TFclusters,haynes99}), but includes a significant amount of new data, plus reprocessing of the old values.  Since SFI++ is a very large and homogeneous sample of nearby spiral galaxies, it is perfectly suited for the study of scaling relations in disc galaxies.  An I-band Tully-Fisher template relation \citep{tf77} is presented in detail for the SCI subsample in \citet{TFclusters}, and for the larger SFI++ cluster sample in \citet[][thereafter M06]{masters06}. Here, the values of half-light radius, absolute magnitude and surface brightness are used as calibration for high-$z$ data.

\subsection{Disc Scaling Relations}

Under the hierarchical scenario for the growth of structure, scaling relations between these three quantities and the rotation velocity of dark matter systems is predicted \citep{mo98}.  The most commonly used of these, which is also the one with the less scatter, is the magnitude-rotation velocity (Tully-Fisher) relation .

\subsubsection{The Magnitude-Rotation Velocity Relation}

\begin{figure}
\resizebox{\hsize}{!}{\includegraphics{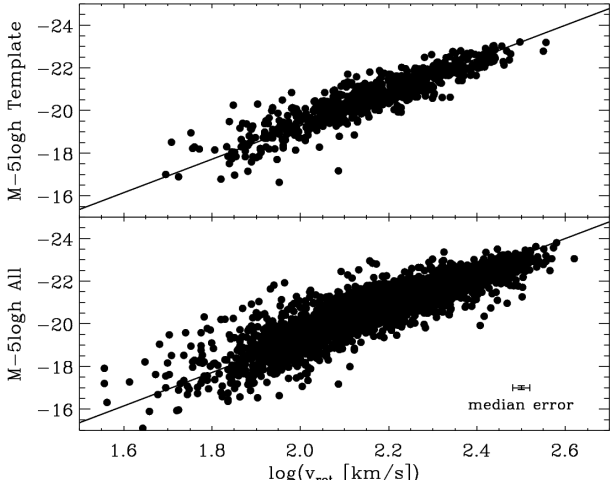}}
\caption{
Magnitude-rotation velocity relation for the sample of $\sim 4200$ SFI++ galaxies.  Absolute magnitude at I-band, corrected for extinction and corrected to face-on \citep{spr07}, plotted against rotation velocity. The solid line in each panel is the template Tully-Fisher relation derived by M06, using only the subset of galaxies with the most accurate distances (top panel). The bottom panel includes all the SFI++ galaxies.}
\label{mag_vrot}
\end{figure}

The Tully-Fisher relation for the SFI++ sample is shown in Figure \ref{mag_vrot}. The best fit to the Tully-Fisher relation at I-band has been derived by M06 for a late-type subset of these galaxies, for which accurate distances can be obtained through membership in nearby clusters of galaxies. We will from now on refer to this subset of galaxies as the "template sample". The Tully-Fisher relation from M06 is:
\begin{equation}
M-5\log h = -20.85 \pm 0.03 - (7.85 \pm 0.1)(\log v_{rot} + \log 2 -2.5)
\label{tf_eq}
\end{equation}
We adopt this expression as our magnitude-velocity relation. M06 also give a precise estimate of the scatter around that relation. We recompute it here however for the complete SFI++ sample, to account for the broader morphological types included and the additional scatter coming from generally larger distance uncertainties. We measure the scatter to be $\sigma=0.56$ mag over the entire range of rotation velocities. 


\subsubsection{The Size-Rotation Velocity Relation}

\begin{figure}
\resizebox{\hsize}{!}{\includegraphics{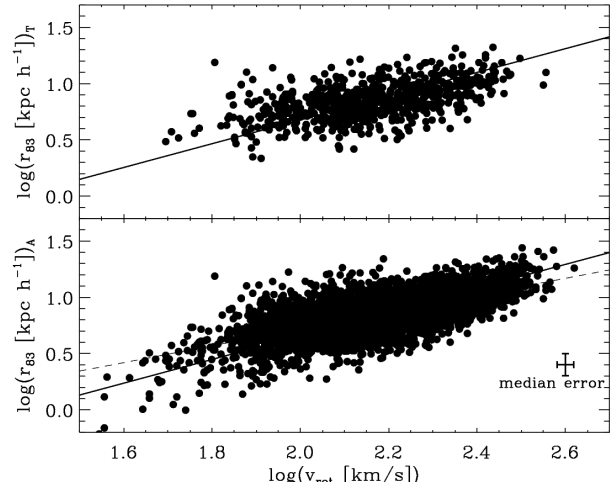}}
\caption{
$r-v_{rot}$ relation for the template sample (top panel) and for the full SFI++ sample (bottom panel). The quantity plotted is $r_{83}$, the radius encompassing 83\% of the light. No attempt is made to correct $r_{83}$ for inclination. In each case the solid line is the best bisector linear fit to the relation, the parameters of which are given in Table \ref{params_fits}.}
\label{r83_vrot}
\end{figure}

\begin{figure}
\resizebox{\hsize}{!}{\includegraphics{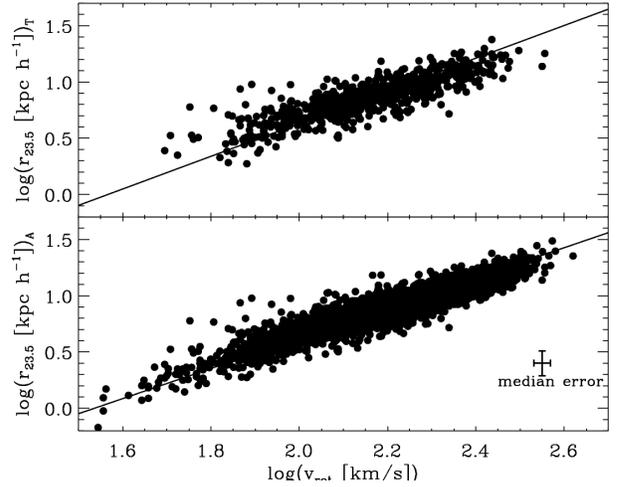}}
\caption{
$r-v_{rot}$ relation for the template sample (top panel) and for the full SFI++ sample (bottom panel). The quantity plotted is $r_{23.5}$, the radius corresponding to the $\mu=23.5$ isophote, after corrections for inclination, foreground extinction and cosmological stretch. In each case the solid line is the best bisector linear fit to the relation, the parameters of which are given in Table \ref{params_fits}.}
\label{r235_vrot}
\end{figure}

Since one of the main goals of this series of papers is to perform the angular diameter test on a sample of high redshift galaxies by using the relation between the physical size of a disc and its rotation velocity to select standard rods, it is most important to derive a reliable expression for the $z=0$ size-rotation velocity ($r-v_{rot}$) relation. \citet{spr07} published two sets of sizes for the galaxies in the SFI++ sample: $r_{83}$, the radius containing 83\% of the light, and $r_{23.5}$ the radius corresponding to the $\mu=23.5$ isophote. While values of the disc scale length $r_d$ are available for SFI++ galaxies, they tend to be unreliable and produce scaling relations with large scatter \citep{giovanelli94}. In Figures \ref{r83_vrot} and \ref{r235_vrot} we present the $r-v_{rot}$ for $r_{83}$ and $r_{23.5}$, respectively.  The scatter around the best-fit bivariate linear fit is smallest when $r_{23.5}$ is used. The Pearson correlation coefficient of the relation is 0.92 for $r_{23.5}$ and 0.67 for $r_{83}$. We will therefore not consider $r_{83}$ any further, but the parameters of the best direct, inverse and bivariate linear fits \citep{isobe} to the $r_{83}-v_{rot}$ relation are given in Table \ref{params_fits}, as well as those for $r_{23.5}-v_{rot}$. 

The smaller scatter in the $r_{23.5}-v_{rot}$ relation is explained by the fact that $r_{23.5}$ as opposed to $r_{83}$ is measured where the disc is close to optically thin, so it is possible to correct it for inclination effects.  While no inclination correction has been attempted for $r_{83}$, the values of $r_{23.5}$ plotted in Figure \ref{r235_vrot} have been corrected for inclination to their face-on values using Equation 7 of \citet{giovanelli95}, where a complete analysis of the inclination dependence of isophotal radii was done. We have further corrected these radii for foreground galactic extinction as:
\begin{equation}
r_{23.5}=r_{23.5}^{o}+\frac{A_i r_d}{1.086},
\end{equation}
where $r_{23.5}^{o}$ is the observed value corrected to face-on, $A_i$ the galactic extinction at I-band derived from the {\it DIRBE} dust maps \citep{schlegel}, and $r_d$ the disc scale length. Both the corrections for inclination and extinction require a measurement of $r_d$, but this quantity was not published by \citet{spr07} for the SFI++ sample, because less reliable than the isophotal measurements of radii. We use these unpublished values of $r_d$ to derive as a function of magnitude the ratio  $r_{23.5}/r_d$, which is then used to scale the reliably-measured values of $r_{23.5}$ down to $r_d$ (this process follows the study of \citet{giovanelli95}). The small scatter in the $r_{23.5}-v_{rot}$ can also be partly explained by the care that has been taken in deriving distances for the SFI++ galaxies. These distances were calculated by taking into account the peculiar velocity of each galaxy. For field galaxies, the peculiar velocities are derived from the systemic velocity of the galaxies and the absolute magnitude compared to the values predicted by the Tully-Fisher relation.  For galaxies in groups, the peculiar velocity is the weighted average of the peculiar velocities of the group members \citep{spr07}. The distances are further corrected for the Malmquist bias.

\begin{figure}
\resizebox{\hsize}{!}{\includegraphics{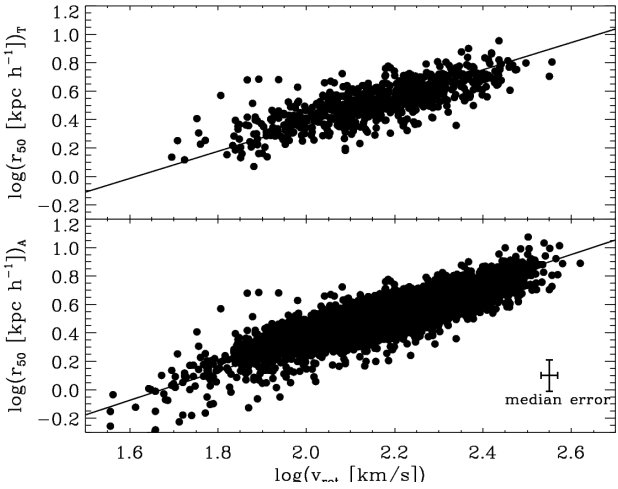}}
\caption{
$r-v_{rot}$ relation for the template sample (top panel) and for the full SFI++ sample (bottom panel). The quantity plotted is $r_{50}$, which has been computed as $r_{50}=1.68r_d$, where $r_d$ the disc scale length was estimated by scaling the value of $r_{23.5}$ by a factor depending on the magnitude of the galaxies. In each case the solid line is the best bisector linear fit to the relation, the parameters of which are given in Table \ref{params_fits}.}
\label{r50_vrot}
\end{figure}

To calibrate the geometrical tests performed in Paper III, the relation between the half-light radius $r_{50}$ and rotation velocity needs to be derived. While it would be straightforward to calculate $r_{50}$ from $r_{83}$, we rather choose to calculate it from $r_{23.5}$ since the $r_{23.5}-v_{rot}$ relation is significantly tighter.  Assuming that the discs obey an exponential profile, the half-light radius is simply $r_{50}=1.68r_{d}$. Instead of using the directly measured but inaccurate values of $r_d$, we use the values obtained by scaling down the inclination- and extinction-corrected values of $r_{23.5}$ as explained above. The $r_{50}-v_{rot}$ relation is presented in Figure \ref{r50_vrot}, and the parameters of the best-fit linear relations are in Table \ref{params_fits}. This  $r_{50}-v_{rot}$ relation is central to the geometrical tests of cosmological models performed in Paper III.

\begin{table*}
\caption{Coefficients of best direct, inverse and bivariate linear fits to the $r-v_{rot}$ relation, $\log(r {\rm [kpc~h^{-1}]})=a+b\log(v_{rot} {\rm [km~s^{-1}]})$.}
\label{params_fits}
\centering
\begin{tabular}{lccccccc}
\hline\hline
Relation & $a_{dir}$ & $b_{dir}$ &  $a_{inv}$ & $b_{inv}$ & $a_{bi}$ & $b_{bi}$ & $\sigma_{bi} (\log(r {\rm [kpc~h^{-1}]}))$\\
\hline
Template galaxies & & & & & & & \\
$r_{83}-v_{rot}$ & $-0.36\pm0.08$ & $0.56\pm0.04$ & $-3.56\pm0.29$ & $2.04\pm0.13$ & $-1.43\pm0.06$ & $1.06\pm0.03$ & 0.16 \\
$r_{23.5}-v_{rot}$ & $-1.35\pm0.06$ & $1.03\pm0.03$ & $-2.27\pm0.07$ & $1.45\pm0.03$ & $-1.77\pm0.06$ & $1.22\pm0.03$ & 0.10 \\
$r_{50}-v_{rot}$ & $-1.06\pm0.05$ & $0.73\pm0.02$ & $-2.16\pm0.09$ & $1.24\pm0.04$ & $-1.54\pm0.04$ & $0.96\pm0.02$ & 0.11 \\
\hline
All SFI++ & & & & & & & \\
$r_{83}-v_{rot}$ & $-0.78\pm0.03$ & $1.05\pm0.01$ & $-2.12\pm0.02$ & $1.06\pm0.01$ & $-1.45\pm0.07$ & $1.05\pm0.03$ & 0.14 \\
$r_{23.5}-v_{rot}$ & $-1.64\pm0.02$ & $1.15\pm0.01$ & $-2.06\pm0.03$ & $1.34\pm0.01$ & $-1.84\pm0.02$ & $1.24\pm0.01$ & 0.07 \\
$r_{50}-v_{rot}$ & $-1.39\pm0.02$ & $0.88\pm0.01$ & $-2.09\pm0.04$ & $1.20\pm0.02$ & $-1.71\pm0.02$ & $1.02\pm0.01$ & 0.08 \\
\hline
\end{tabular}
\end{table*}

\subsubsection{The Surface Brightness-Rotation Velocity Relation}

\begin{figure}
\resizebox{\hsize}{!}{\includegraphics{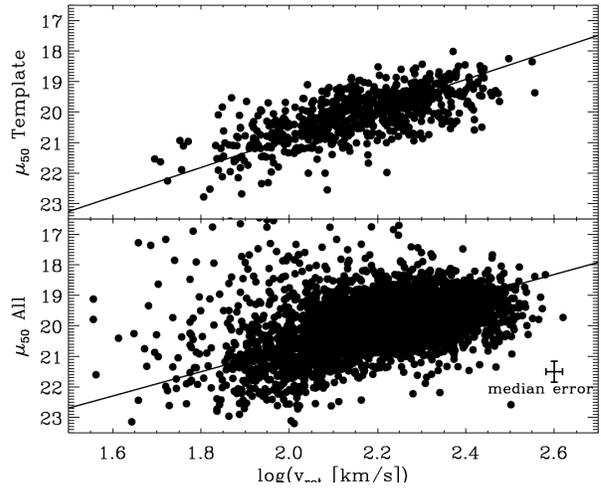}}
\caption{
Surface brightness-rotation velocity relation for the sample of $\sim 4200$ SFI++ galaxies.  Surface brightness at I-band within $r_{50}$, the radius containing 50\% of the light, plotted against rotation velocity. The top panel uses only the smaller sample used by M06 to derive the template Tully-Fisher relation, while the bottom panel includes all the SFI++ galaxies.}
\label{sb_vrot}
\end{figure}

The surface brightness-rotation velocity ($\mu-v_{rot}$) relation is the least often used of these three scaling relations, as it intrinsically has more scatter than both the Tully-Fisher and the $r-v_{rot}$ relations. This larger scatter is due to a stronger dependency of the $\mu-v_{rot}$ relation on the spin parameter of the disc $\lambda$; while the $M-v_{rot}$ relation is independent of $\lambda$, this parameter appears linearly in the $r-v_{rot}$ relation and quadratically in the $\mu-v_{rot}$ relation \citep{mo98}.

In Figure \ref{sb_vrot} we show the $\mu-v_{rot}$ relation both for the template sample and for the entire SFI++ relation.  The quantity plotted is the average surface brightness within $r_{50}$, calculated as:
\begin{equation}
\mu_{50}=m_{tot} + 5\log (r_{50}) + 2.5\log(2\pi).
\end{equation}
The best fit relation to the entire SFI++ sample, using an ordinary least squares bisector method, is:
\begin{equation}
\mu_{50}=(29.67 \pm 0.22) - (4.43 \pm 0.09) \log(v_{rot})
\end{equation}
The scatter of the residuals of this best fit relation is $\sigma=0.78$.

\section{CONCLUSIONS \label{discussion}}

The main goal of this paper was to provide calibration in two different ways for the angular diameter test that we performed on data from the VVDS: (1) by directly comparing velocity indicators commonly used in the local and distant Universe, and (2) by using a large sample of nearby galaxies to provide zero points for the radius - rotation velocity, magnitude - rotation velocity and surface brightness - rotation velocity relations.  Previous studies have focused on comparing the H${\alpha}$ or [OII] rotation velocities with HI data for nearby galaxies.  \citet[]{kobulnicky} have used a sample of 22 nearby galaxies to compare [OII] and HI linewidths.  Their galaxy sample was smaller and less homogeneous that the one used in this study.  They found the two sets of velocity estimates to be consistent within $10\%$ overall, with the [OII] underestimating the HI rotation velocity by up to $50\%$ in a few extreme cases.  In the local Universe, the consistency of the HI and H${\alpha}$ rotation velocities was also established \citep[see for example][]{courteau97,vogt}.  Most recently, \citet[]{catinella} has shown that on average HI widths are larger than H${\alpha}$ ones by $\sim10\%$ for a galaxy with $w=100$ km s$^{-1}$ and that optical rotational widths measured from velocity histograms are affected by systematic biases and therefore are less reliable that those derived by using the full spatial information through the fitting of the rotation curves. 

For this study, we first analysed the rotation curves for all the galaxies in our sample and concluded that the rotation velocities derived from the [OII]$\lambda 3727$ \AA\ line translate directly into the velocities obtained from H${\alpha}$ rotation curves, given that the sampling of the [OII] rotation curve extends far enough to reach the region of constant (or slowly increasing) velocity.  Even under this condition and as shown in Figure \ref{vrot_plot}, there is scatter in the relation of about 30 km s$^{-1}$.  The next step was to compare the techniques used to determine rotation velocities in the local Universe (H${\alpha}$ rotation curves) and at high redshift ([OII] velocity histograms).  Once again, there is a direct relation between the two but significant scatter on the order of 50 km s$^{-1}$ (see Figure \ref{sigma_plot}).  

We conclude that all the rotation velocity indicators studied here give comparable results, but before combining samples analysed with different methods one should be aware of the important scatter in the different relations.  This scatter is due mostly to the poorer spatial extent of the [OII] rotation profiles compared to the H${\alpha}$ ones obtained with the same integration time, and can be as large as $25\%$ for a galaxy with $w=200$ km s$^{-1}$ or $100\%$ if $w=50$ km s$^{-1}$ in the case of the $\sigma_{rot}$([OII])-$v_{rot}$(H${\alpha}$) relation (Fig.\ref{sigma_plot}).  Even considering this large scatter, the fact that there is no systematic deviation from a 1:1 relation allows us to use both methods in their respective context to divide galaxies in classes based on rotation velocity, as required to perform the geometrical tests with the high redshift sample (see Paper III).  In Paper I it was showed that this is a viable way of selecting standard rods for the purpose of the angular diameter test.  Since one cannot perform the test only with high redshift data, it was of great importance to establish that a set of local calibrators can be reliably compared to it, even if the rotation measure is different.

Once it is established that the various rotation indicators can be reliably compared, the next critical step is to derive the scaling relations between luminosity, size, surface brightness and rotation velocity for a set of nearby galaxies, free from any evolution. We used the SFI++ catalog, a very large homogeneous sample of nearby spiral galaxies with I-band photometry and rotation information obtained either through optical long-slit spectroscopy or HI measurements. We have particularly focussed our attention on the $r-v_{rot}$ relation which is central to the cosmological tests performed in the other papers of this series. We have shown that by carefully correcting sizes for inclination and extinction effects as well as obtaining accurate distances to the galaxies, the $r_{23.5}-v_{rot}$ relation can be almost as tight as the Tully-Fisher relation, as judged by their Pearson correlation coefficients.  The relation derived using $r_{23.5}$ as the size measurement is considerably tighter than the ones obtained using the disc scale length $r_d$ or a radius defined to contain a given fraction of the light of the galaxy (in our case $r_{83}$ or $r_{50}$).  The next phase for this study on scaling relations will be to see what constraints our $r-v_{rot}$ relation imposes on models for disc galaxy formation, for example the range of spin parameters its tight scatter allows. These scaling relations also offer exciting possibilities for the study of disc galaxy evolution. Since they provide accurate values for luminosity, size and surface brightness as a function of rotation velocity, measurements of the properties for galaxies with the same rotation velocity (i.e. hosted in a dark halo of the same mass) over a large redshift range can provide in a new and unique way information about the evolution of galaxies and a mean of testing predictions from models under the hierarchical scenario for the growth of structures. A pilot study is presented in the next paper of this series.

\begin{acknowledgements}
This work has been partially supported by NSF grants AST-0307661 and AST-0307396 and was done while AS 
was receiving a fellowship from the {\it Fonds de recherche sur la Nature et les Technologies du Qu\'{e}bec}.  
KLM is supported by the NSF grant AST-0406906.
This research has made use of the NASA/IPAC Extragalactic Database (NED) which is operated by the 
Jet Propulsion Laboratory, California Institute of Technology, under contract with the National Aeronautics 
and Space Administration.
\end{acknowledgements}

\end{document}